**Ali R. Baghirzade**
Master of economics
Junior researcher of the research Institute "Innovative Economics"
Plekhanov Russian University of Economics Moscow, Russia
e-mail: Bagirzade.AR@rea.ru

# MODERN RISKS OF SMALL BUSINESSES

**Abstract**
An important area of anti-crisis public administration is the development of small businesses. They are an important part of the economy of developed and developing countries, provide employment for a significant part of the population and tax revenues to budgets, and contribute to increased competition and the development of entrepreneurial abilities of citizens. Therefore, the primary task of the state Federal and regional policy is to reduce administrative barriers and risks, time and resources spent on opening and developing small businesses, problems with small businesses ' access to Bank capital [8], etc. Despite the loud statements of officials, administrative barriers to the development of small businesses in trade and public catering are constantly increasing, including during the 2014-2016 crisis.

Until 2010, the most common type of small business was non-stationary objects of small-scale trade - car counters, trays, tents, etc. In accordance with the decree of the government of Moscow of January 27, 2004 No. 29-PP "on regular measures to streamline the operation and placement of small-scale retail facilities in the territory of Moscow", they are established without registration of land and legal relations, i.e. without guarantees of long-term operation. For these objects, the permit document for the right to trade is "permission to place an object". The absence of a lease agreement for a land plot under a non-stationary object meant that it could be moved to another less passable place at any time or completely closed.

Prior to 2007, the process of obtaining a permit was quite simple. A permit for the right to trade was issued by the district Council in the presence of a package of permits. This package included: the decision of the consumer market Commission of the Prefecture of the district, approval with the Head of the district Administration, conclusions of Rospotrebnadzor, gosnadzor, ATI, permission to connect electricity, registration of the cash register in the Federal tax service, and also subject to the inclusion of this outlet in the layout of non-stationary small-scale retail facilities in the district. The process of obtaining permits on average did not exceed three months, and the cost of its implementation was 180 thousand rubles.

But since 2007, in accordance with the decree of the Moscow Government No. 274 - PP of 25.04.2006, the allocation of places for non-stationary small-scale retail facilities has been carried out only on a competitive basis. To get a permit to place an object for up to one year, you now need to collect the entire package of permits, and then win the competition. As a result, the situation of non-stationary objects of small-scale trade has become even more unstable, the time for issuing permits has increased to 6 months, as well as the cost of obtaining them to 250 thousand rubles. Despite these measures, in 2010 more than 20 thousand small - scale retail objects were installed in Moscow-kiosks, tents and car stands (tonars).

In addition, until 2008, the most attractive type of small business in Moscow was the installation of non-capital objects - bus stops and shopping modules and shopping pavilions, since they were allowed to issue a land lease agreement for up to five years with the possibility of extension. During the lease period, you didn't have to worry about the fate of your business: you can only close the company by agreement of the parties (if you provide an equivalent land plot) or by a court decision. Entrepreneurs were not deterred by a significant increase in the time required for processing documentation, nor by a significant increase in the cost of its implementation. It was necessary to obtain the decision of the Interdepartmental Commission on consumer market issues under the government of Moscow, the decisions of the Commission on trade and the land Commission of the Prefecture of the administrative district, and the approval of the district Council. As well as develop a situation plan and town-planning conclusion, get approval from Rospotrebnadzor, uspn of the Ministry of emergency situations, the Department of nature management of Moscow, state unitary enterprise Mosgorgeotrest, issue and register a land lease agreement. Prior to 2008, the territorial departments Of the Moscow Department of land resources issued not only lease agreements for land plots, but also carried out all the necessary geodetic measurements, as well as put the plots on cadastral registration. Therefore, the process of registration of land relations from choosing a place to registering a lease agreement took an average of 1.5 years, and the cost of registration of one place was an average of 500 thousand rubles. But since 2008, the territorial departments Of the Department of land resources have ceased to be engaged in geodesy and cadastral registration of land plots and have transferred these functions to commercial structures. As a result, the terms of registration of land relations stretched to 2 years, and the additional costs of registration increased to an average of 650 thousand rubles. The process of drawing up lease agreements was costly and time-consuming, but generally acceptable for Metropolitan entrepreneurs who opened more than 2 thousand stop-and-trade modules and 1 thousand pavilions in Moscow.

However, with the change of leadership in the Moscow Government, the situation has changed dramatically - the process of cleaning up the streets of the capital from small businesses and, above all, retail and public catering enterprises has begun. In 2010-2015, all bus stops and shopping centers, 13 thousand tents, kiosks and pavilions were dismantled in Moscow. All that remained were newsstands, ice cream parlors, and theater tickets.

Yet the decisive year "cleansing" of the city from small and medium-sized businesses can call 2016 the Legal basis for further action to accelerate the demolition of commercial facilities was the Moscow government Resolution of December 8, 2015 N 829-PP "On measures to ensure the demolition of unauthorized construction in some areas of the city of Moscow", approved due to the change in article 222 of the Civil code. The new version of the article of the civil code of the

Russian Federation allowed state bodies to demolish unauthorized buildings if they are located in areas with special conditions (near utility lines, etc.) without a court decision.

Following this resolution, the first part of the list of "unauthorized outposts" was published, which included 104 shopping facilities, including shopping malls near the Chistye Prudy metro station, kiosks on Kropotkinskaya street, a building above the exit from the Sukharevskaya metro station, the Pyramid shopping center on Pushkinskaya square, tents on Mayakovskaya, Sokol, Novoslobodskaya, etc.

Until recently, the owners of the largest objects did not believe that this threat was real, so they were not going to demolish the sources of income. They pointed out the contradictions that have arisen: according to the law, only the object, the right of ownership to which is not confirmed by the court and is not registered in the Unified state register of rights to immovable property, is considered to be a self-construction project, and these documents were issued by most of them [1, p. 23]. They also pointed out that resolution 829-PP contradicted Federal law No. 122-FZ of 21.07.1997 "on state registration of rights to immovable property and transactions with it", article 2 of which States that "a registered right to immovable property can only be challenged in court". They also referred to article 35 of the Constitution, which sets out the constitutional principle of inviolability of property. But as a result, the owners of shopping facilities were not heard and on the night of February 9 this year, dozens of shopping kiosks and pavilions were simultaneously dismantled in the city. The courts were flooded with lawsuits from businessmen, many of whom provided evidence that the construction of the facility was authorized by the city authorities and that all permits are available [2, p. 7]. However, the courts sided with the Moscow authorities. These include the decision of the constitutional court of September 27, 2016, which confirmed the constitutionality of the dismantling mechanism prescribed in the Civil code. "The mechanism used in Moscow does not violate the principles laid down in the Constitution of the Russian Federation, but, on the contrary, protects public interests, helps to create and maintain a comfortable living environment for citizens, protect their lives and health," the press service of the mayor and government of Moscow told the essence of the constitutional court decision.

The mass demolition of pavilions and kiosks, popularly known as the "night of long buckets", caused a great response. Citizens were outraged that now they have no place to buy a loaf of bread on the way home, and for a bottle of water they have to stand in queues at the supermarket. Entrepreneurs complained about how they lost their last source of income. But the Metropolitan authorities were ignored.

The second phase of demolition was completed in late summer 2016. At that time, 107 objects were supposed to be dismantled, including large shopping centers in residential areas and shopping malls near suburban metro stations. However, by the time of the second demolition, the city authorities had already managed to prescribe a compensation mechanism, and businessmen realized that the judicial prospects were hopeless. Therefore, out of 107 objects, 77 owners immediately gave permission for voluntary dismantling.

Then the Moscow Government published the third list of particularly dangerous objects of unauthorized construction. It includes 43 objects with a total area of 26.7 thousand square meters, most of which (15 units, 21.2 thousand square meters) are located in the Northern administrative district. Among other things, two large shopping centers in the area are being demolished at once: the Petrovsko-Razumovsky shopping center and the Europa shopping center. In the center of the capital, a two-story house will also be demolished at 57/25 Bakuninskaya street; several one-story shops with food and flowers at Zemlyany Val, Krasnoprudnaya street, Bolshaya Polyanka, Shchepkin street and Sergiy Radonezh street. The three-story shopping center on Kashirskoe shosse 27, the two-story shopping center on Novocherkasskiy Boulevard, and two long rows of tents near the Medvedkovo and Vykhino metro stations will disappear. Not only small businesses, but also large networks are affected. In particular, in the listed shopping facilities there are two stores "Magnolia" (on Lyusinovskaya street and Marshal Zhukov Avenue) and one "Pyaterochka" (on Isakovsky street), salons of mobile operators MTS, MegaFon, Beeline.

Most items are considered dangerous because they are located in protected areas of utility lines or linear objects (for example, metro stations). But there are several points that are close to natural areas. For example, a building with a car service station located on the border with Izmailovsky forest Park and a huge shopping center near Stroginskaya floodplain (6 Isakovsky street) are subject to demolition.

Owners are given two months to dismantle buildings voluntarily: on their own or by contacting the help of the Prefecture. In this case, the owners can count on compensation: 55.5 thousand rubles for each demolished "square" in case of independent demolition and 51 thousand rubles if the forces of the Prefecture were involved. They are also promised assistance in selecting new areas for relocation. If the owners refuse to dismantle the objects, they will be forcibly demolished in two months.

It is not yet known how many objects that may be considered dangerous self-construction in the future (after the third wave of demolitions). But this work continues: the capital's state real estate Inspectorate confirmed that " new objects will be identified as the technical documentation is studied."

The current wave of demolitions, like the previous ones, undermines business confidence in the Executive and judicial authorities, reduces entrepreneurial activity, leads to an increase in rental rates, unemployment, and a decrease in the income of residents of the capital and the Moscow region.

**Acknowledgment**

This article was prepared as part of the government contract as requested by the Ministry of Science and Higher Education of the Russian Federation on the subject formulated as «Structural changes in economy and society as a result of achieving the target indicators of National projects, which provide opportunities to organize new areas of social and economic activity, including commercial, both in Russia and abroad» (project No. FSSW-2020-0010)